\documentclass[5p]{elsarticle}

\usepackage{amsmath,amssymb,amsfonts}
\usepackage[official]{eurosym}
\usepackage{algorithmic}
\usepackage{algorithm}
\usepackage{multirow}
\usepackage{stfloats}
\usepackage{subfig}
\usepackage{url}
\usepackage{hyperref}
\hypersetup{
    colorlinks=true,
    linkcolor=blue,
    filecolor=blue,      
    urlcolor=blue,
    citecolor=blue,
    }
\usepackage{textcomp}

\journal{}

\begin{document}

\begin{frontmatter}

\title{The Impact of PV Generation Forecast and Multi-Objective Control Policy on Optimal Operation of Grid Connected PV-BESS Microgrid}

\author[inst1]{Berhane Darsene Dimd\corref{mycorrespondingauthor}}
\ead{berhane.d.dimd@ntnu.no}
\cortext[mycorrespondingauthor]{Corresponding author}

\affiliation[inst1]{organization={Department of Electric Energy},%Department and Organization
            addressline={Norwegian University of Science and Technology (NTNU)}, 
            city={Trondheim},
            postcode={7491}, 
            country={Norway}}

\author[inst1]{Steve Völler}
\ead{steve.voller@ntnu.no}
\author[inst1]{Ole-Morten Midtgård}
\ead{ole-morten.midtgard@ntnu.no}

\begin{abstract}
%% Text of abstract
The variability of photovoltaic (PV) generation poses significant challenges to the reliable and efficient operation of grid-connected microgrids. Accurate PV output power forecasting and efficient energy scheduling strategies are essential not only for optimizing PV system operation but also for improving the overall performance and reliability of the system. This study proposes a long short-term memory (LSTM)-based PV power forecasting model integrated with a multi-objective scheduling framework for a grid-connected PV–battery energy storage system (BESS). The proposed approach enables detailed performance monitoring and assessment by quantifying how PV forecast accuracy influences key operational metrics, including PV self-consumption ratio, grid energy cost, grid injection, and battery utilization. Three forecasting scenarios (perfect forecast, persistence model, and LSTM-based forecast) are compared to evaluate their impact on system performance and operational reliability. Results show that the LSTM-based forecast reduces root mean squared error (RMSE) by 6\% compared with the persistence model, increases the PV self-consumption ratio from 78.1\% to 84.5\%, and reduces grid injections by 82\%. The analysis also highlights trade-offs, as higher battery throughput associated with improved performance may contribute to accelerated aging. These findings demonstrate the importance of accurate PV forecasting in improving system performance and ensuring reliable operation. Future work will focus on probabilistic forecasting to properly quantify uncertainties, incorporate load prediction, and develop smart control strategies that allow grid-support functionalities from the PV side.

\end{abstract}


\end{frontmatter}

%% main text
\section{Introduction}
% \linenumbers
The achievement of the United Nations' (UN) ambitious energy and climate targets by 2030 necessitates a concerted effort towards the development and implementation of renewable energy resources (RER). The global contribution of RER in electricity generation is forecast to rise from 30\% in 2023 to 37\% in 2026, with the growth largely supported by the expansion of solar photovoltaic (PV) generation \cite{IEA2024electricity}. However, these advancements are not enough to the scale required to fulfill the UN's Sustainable Development Goals, particularly Goal 7, which emphasizes the importance of affordable, clean and sustainable energy. Motivated by this imperative and driven by various economic factors, the global revolution in PV systems has gained momentum. PV electricity, even in regions rich in hydro and wind power resources, is emerging as a compelling alternative. Nevertheless, realizing the full potential of PV electricity requires addressing the inherent challenges since it will be operated mostly in combination with the conventional grid.

A key challenge in grid-connected PV systems lies in their sensitivity to environmental conditions, resulting in uncertain and irregular output power. This inherent variability impacts the fundamental operational objective of balancing variable loads with variable supply. To overcome this challenge, an battery energy storage system (BESS), a PV output power forecasting system, and an energy scheduling (ES) system can be employed in microgrid operation mode. Microgrids, characterized by their localized generation and consumption of electricity, facilitate this optimization \cite{adefarati2019reliability}. Such a combined system enables the optimization of power exchange within the microgrid and with the main grid. However, the effectiveness of such an approach is dependent upon the accuracy of the PV output power forecast model and the implemented control policy within the ES system \cite{gandhi2024value}.

The conventional ES policies promote charging the BESS whenever there is excess PV energy in the system \cite{bertsch2017drives, hassan2017optimal}. This policy ensures that the largest portion of the PV energy is consumed locally. However, this approach has a negative impact on both the BESS and the grid. Frequent charging/discharging of the BESS leads to battery degradation due to cycling aging \cite{xu2016modeling, alam2016cycle}. Furthermore, early charging may cause the peak generation from the PV system to coincide with low peak electricity demand, typically occurring around solar noon. This results in a large surge of PV energy into the grid during this period, which can result in grid congestion and potentially lead to voltage and frequency fluctuations in the grid \cite{tevar2019influence, dhlamini2018solar, rahman2021analysis}. Consequently, grid operators will be compelled to invest in adequate infrastructure and control mechanisms (such as PV curtailment) to maintain grid stability. This is not an economically viable solution.

 \begin{table}[!ht]
\small
\begin{tabular}{|p{1.3cm}p{6.7cm}|}
\hline
   &\\
\multicolumn{2}{|l|}{\textbf{Nomenclature}} \\
 & \\
\multicolumn{2}{|l|}{\textbf{Abbreviations}} \\
BESS & Battery Energy Storage System \\
ES & Energy Scheduling\\
PV & Photovoltaic\\
MPC & Model Predictive Control\\
MILP & Mixed Integer Linear Programming\\
LSTM & Long Short-Term Memory\\
 & \\
\multicolumn{2}{|l|}{\textbf{Battery Parameters and Variables}} \\
$x_k$ & State of charge of the BESS \\
$\eta_{ch}, \eta_{dis}$ & Charging and discharging efficiency of the BESS\\
$SOC_{min}$ & Minimum state of charge of the BESS\\
$SOC_{max}$ & Maximum state of charge of the BESS\\
$P_{min}^b$ & Minimum charging/discharging power\\
$P_{max}^b$ & Maximum charging/discharging power\\
 & \\
\multicolumn{2}{|l|}{\textbf{Grid Parameters and Variables}}\\
$\gamma_{k}^{pur}$ & Prices for buying energy from the grid\\
$\gamma_{k}^{sell}$ & Prices for selling energy to the grid \\
$P_{min}^g$ & Minimum power exchange with grid \\
$P_{max}^g$ & Maximum power exchange with grid \\
 & \\
\multicolumn{2}{|l|}{\textbf{Decision Variables}}\\
$P_k^b$ & Power exchange with the BESS at $k$ in $kW$ \\
$\delta_k^b$ & Binary variable for the BESS. Equals to 1 if discharing, 0 otherwise\\
$\psi_k^b$ & Auxiliary variable for the BESS \\
$P_k^g$ & Power exchange with the grid at $k$ in $kW$ \\
$\delta_k^g$ & Binary variable for the grid. Equals to 1 if buying from the grid, 0 otherwise \\
$\psi_k^g$ & Auxiliary variable for the grid \\
 & \\
\multicolumn{2}{|l|}{\textbf{Other Parameters}}\\
$N_p$ & Prediction horizon \\
$\Delta t$ & Discrete time step \\
$k$ & Discrete time step index\\
$P_k^{pv}$ & The power injected by the PV system at $k$ \\
$P_k^{l}$ & The power consumed by the load at $k$ \\
$\alpha_{cyc}$ & Battery cycling cost\\
$\lambda_{imp}$ & Grid import penality\\
\hline
\end{tabular}
\end{table}

Recent years have seen a rise in the number of ES techniques formulated as multi-objective optimization problems, encompassing diverse cost functions and parameters, including operational costs, storage degradation costs, and spot market rates \cite{bordons2020basic}. They use model predictive control (MPC) as a core control technique to compute actions that optimally satisfy these objectives. The precise implementation details of MPC for ES problem differ based on various factors, such as problem formulation, cost functions, prediction horizon, grid interaction, provision of ancillary services, and other considerations.

PV output power forecasting is a key element in the implementation of a multi-objective ES policy \cite{gandhi2024value}. It allows for precise estimation of available PV energy and enables well-informed decisions concerning optimal power dispatch, thus improving economic efficiency. This forecasting capability enables operators to avoid unnecessary grid imports and injections, as well as excessive generation capacity, thereby curtailing operational costs and optimizing the utilization of renewable energy resources \cite{hao2020power, nwulu2017optimal}.

An overview of related literature is presented below. In the work of \cite{nair2021analysis, deng2022economic}, MPC is used to mitigate grid congestion and battery degradation while maximizing the local consumption of PV energy. Their study demonstrated the effectiveness of MPC-based ES under PV generation forecast uncertainties. Despite the success of these works in many aspects, these studies did not consider the energy price from the grid to optimize the system. In contrast, a study that considers grid tariffs in the optimization problem is presented in \cite{bordons2020energy}. The operation of the grid is optimized in such a way that the purchase of energy from the grid is done only when the prices are lower, and energy sale to the grid is done at maximum price periods. Unlike the former works that included grid issues in the optimization problem, this work only focuses on costs associated with BESS degradation and operation.

Several other studies have explored the economic scheduling of BESS-based microgrids using MPC, as evidenced by the works of \cite{nunez2017optimal, garcia2015optimal, conte2020economic}. However, the predominant focus of these studies is on the ES policy aspect, often neglecting or providing minimal attention to the PV generation forecast aspect. In conclusion, while acknowledging the valuable contributions of the reviewed papers, it is observed that they lack to varying extents one or more of the following crucial aspects of ES based on MPC control:
    \begin{itemize}
        \item The BESS and PV system model utilized in the study exhibited a simplified nature, lacking the necessary complexity needed for fair representation of real-world dynamics.
        \item The study did not sufficiently address the challenges posed by uncertain and irregular PV output power through the use of a PV output forecast model.
        \item The impact of the ES policy on the grid was not adequately addressed in the study, overlooking an essential aspect of system integration. 
        \item The study did not incorporate real-time pricing information from the grid side to optimize the operation of the microgrid. 
    \end{itemize}
The above observations underscore the necessity for further research and exploration in this area.

This study, therefore, proposes a holistic ES framework based on PV generation forecasting and multi-objective control policy that takes into consideration both the economic and technical aspects of the system. In tackling this multi-objective optimization problem, the goal is to obtain an optimal solution that represents a trade-off among the competing objectives. To accomplish this, the study proposes a control policy based on Mixed Integer Linear Programming (MILP) and MPC, with the primary aim of achieving the following objectives:

    \begin{enumerate}
    
        \item Cost reduction: The control policy seeks to minimize the cost of buying grid electricity, thereby optimizing the economic efficiency of the system.
        \item Maximize local consumption of PV electricity: The policy aims to maximize the utilization of PV electricity locally, promoting self-consumption and reducing reliance on external grid.
        \item Addressing the impact on the grid: The ES policy focuses on minimizing large injections of PV power into the grid during solar noon when PV generation is typically at its peak.
        \item BESS degradation mitigation: The ES policy aims to reduce degradation of the BESS by penalizing excessive cycling. 
    
    \end{enumerate}

The remainder of this paper is structured as follows. Section \ref{sec:meth} presents the methodology used to address the problem, including a brief description of the microgrid and the formulation of the problem as a mixed-integer linear program. The implementation of MPC for the energy scheduling use case is also discussed in this section. Section \ref{sec:lstm} details the implementation of PV generation forecasting using long short-term memory (LSTM) network. Section \ref{sec:res} presents the main findings and discussion, covering both the PV forecasting results and the optimal operation of the microgrid. Finally, Section \ref{sec:conc} summarizes the key conclusions of this work.

%------------------------------------------------------------------------------------------------------------%
\section{Methodology}
\label{sec:meth}

\begin{figure*}[hb]
        \centering
        \includegraphics[width=\textwidth]{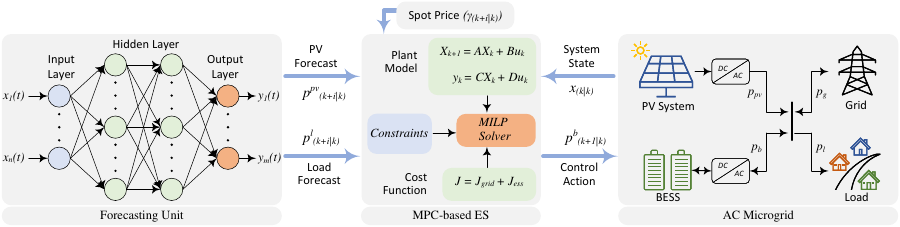}
        \caption{The proposed energy scheduling framework for a grid-connected PV-BESS microgrid. Forecasting system, MPC-based energy scheduling and PV-BESS AC microgrid.}
        \label{fig:micro}
    \end{figure*}

A generic scheme of the MPC-based ES proposed in this work is shown in Figure \ref{fig:micro}. It incorporates a forecast model for PV output power and consumption patterns. The predictions from the forecast model and the cost of electricity from the grid are used as inputs. The ultimate goal of this multi-objective ES is to maintain instantaneous power balance in the microgrid at all times while optimally satisfying various objectives in the control policy. The ES determines the set points of the BESS ($P_{k+i|k}^b$) and the grid ($P_{k+i|k}^g$) based on the current state of the BESS ($x_{k|k}$) such that all objectives are optimally achieved. A day-ahead real-time electricity price is used in this work from Nord Pool \cite{np}.

\subsection{System Description}

This study considers a grid-connected microgrid comprising a PV system, a battery BESS, and a load, as illustrated in Figure \ref{fig:micro}. The microgrid operates in such a way that the PV system supplies the load whenever there is sufficient power from the PV. If there is excess power from the PV system, it is either used to charge the BESS for later use or directly supplied to the grid. Conversely, during periods when the PV system cannot meet the load demand in the microgrid, the BESS or the grid must supply the load. The load in the microgrid is considered to be critical where the demand has to be met at all times.

\subsection{MPC-based Multi-Objective ES}

The ES considered in this work is formulated as a multi-objective control problem where the goal is to obtain an optimum scheduling plan that represents a trade-off among various objectives. For this purpose, MPC, a widely popular control technique, is used. MPC is a feedback control method that utilizes a mathematical model of the system to make predictions about its future behavior and computes control actions based on these predictions \cite{seborg2016process}. MPC is an ideal control technique for multi-variable complex problems, such as the economic operation of a microgrid. 

In the case of microgrid operation, the MPC generates control moves into the future (for a given prediction horizon, $N_p$) that considers the dynamic states and constraints of the microgrid while optimally satisfying multiple objectives. For a given $N_p$ and based on the predictions of the PV power $P_{k+i|k}^{pv}$, the load $P_{k+i|k}^{l}$, the energy spot price $\gamma_{k+i|k}$, and the current state of the system $x_{k|k}$, the MPC computes the future states of the system and control signals (battery power $P_{k+i|k}^b$ and grid power $P_{k+i|k}^g$). The implementation details of MPC for the optimal operation of the microgrid in this work is outlined in Algorithm \ref{alg:mpc}. The various elements within the microgrid are represented through mathematical models in the following manner.

    \begin{algorithm}[tb]
    \caption{MPC algorithm for ES}\label{alg:mpc}
        \begin{algorithmic}
            \STATE 
            \STATE {\textbf{Step 1:}} At sampling instant $k$, read current SOC ($x_{k|k}$) and predicted disturbances ($P_{k+i|k}^{pv}$ and $P_{k+i|k}^{l}$) for $i=1...N_p$.
            \STATE {\textbf{Step 2:}} Compute $N_p$ control moves by solving the optimization problem. i.e., $P_{k+i|k}^b$ and $P_{k+i|k}^g$ for $i=0...N_p-1$.
            \STATE {\textbf{Step 3:}} Implement the first control inputs ($P_{k|k}^b$ and $P_{k|k}^g$) and discard the rest.
            \STATE {\textbf{Step 4:}} Shift the horizon by one step ($k\leftarrow k+1$).
            \STATE {\textbf{Step 5:}} Go to Step 1.
        \end{algorithmic}
    \end{algorithm}

\subsubsection{System State}
For the purpose of this work and the sampling time considered here (i.e. 15 minutes), the system model corresponds to the evolution of the state of charge $x_k$ of the battery. $x_k$, which represents the BESS's level of charge, is calculated using the Coulomb Counting method given by (\ref{eq:1}). The Coulomb Counting method is widely applied for SOC estimation owing to its simplicity, computational efficiency, and suitability for real-time implementation \cite{stefanopoulou2015system}.

    \begin{equation}
        {x_{k}} = \begin{cases}
        x_{k-1} - \frac{\Delta t}{\eta_{dis} \cdot C_b}\cdot P_k^b,&{\text{if}}\ P_k^b \geq 0\\ 
        {x_{k-1} - \frac{\eta_{ch} \cdot \Delta t}{C_b}\cdot P_k^b,}&{\text{otherwise.}} 
        \end{cases}
        \label{eq:1}
    \end{equation}
where $x_k$ is the state of charge of the BESS at instant $k$, $\eta_{ch}$ and $\eta_{dis}$ are the charging and discharging efficiencies, $\Delta t$ is the sampling time, $C_b$ is the battery capacity and $P_k^b$ is the power to/from the battery.

Equation (\ref{eq:1}) is a hybrid model which cannot be used directly in the optimization problem. The piecewise condition in the above equation can be replaced by using a binary variable $\delta_k^b$ such that $P_k^b \geq 0 \Leftrightarrow \delta_k^b = 1$ as described in \cite{bemporad1999control}. This is introduced into the optimization problem as a constraint,
    \begin{subequations}\label{eq:2}
        \begin{align}
            -P_{min}^b \cdot \delta_k^b-P_k^b + P_{min}^b & \leq 0 \label{eq:2A}\\
            -(P_{max}^b + \varepsilon)  \cdot \delta_k^b+P_k^b+\varepsilon &\leq 0 \label{eq:2B}
        \end{align}
    \end{subequations}
where $P_{min}^b$ and $P_{max}^b$ are the minimum and maximum physical limits of the battery power during charging and discharging and $\varepsilon$ is a small positive scalar. Equation (\ref{eq:1}) can now be written as (\ref{eq:22}).
    \begin{equation}
        x_{k} = x_{k-1} - (\frac{1}{\eta_{dis}}-\eta_{ch})\cdot \frac{\Delta t}{C_b}\cdot \delta_k^b \cdot P_k^b - \frac{\eta_{ch}\cdot \Delta t}{C_b}\cdot P_k^b \label{eq:22}
    \end{equation}
The bi-linear term in the above equation ($\delta_k^b \cdot P_k^b$) introduces non-linearity and it can be avoided by defining an auxiliary variable $\psi_k^b$ such that $\psi_k^b=\delta_k^b \cdot P_k^b$. Based on \cite{bemporad1999control}, this auxiliary variable is introduced as a constraint in the optimization problem as in (\ref{eq:3}),

    \begin{subequations}\label{eq:3}
        \begin{align}
            \psi_k^b - P_{max}^b\cdot \delta_k^b &\leq 0 \label{eq:3A}\\
            \psi_k^b - P_{min}^b\cdot \delta_k^b &\geq 0 \label{eq:3B}\\
            \psi_k^b - P_k^b + P_{min}^b(1 - \delta_k^b) &\leq 0 \label{eq:3c}\\
            \psi_k^b - P_k^b + P_{max}^b(1 - \delta_k^b) &\geq 0 \label{eq:3d}
        \end{align}
    \end{subequations}
Therefore, the SOC of the battery can finally be written as (\ref{eq:soc}).
\begin{equation}
  x_{k} = x_{k-1}- (\frac{1}{\eta_{dis}}-\eta_{ch})\cdot \frac{\Delta t}{C_b}\cdot \psi_k^b - \frac{\eta_{ch}\cdot \Delta t}{C_b}\cdot P_k^b
  \label{eq:soc}
\end{equation}

\subsubsection{Interaction with the Grid}
A grid-connected microgrid can sell/purchase energy to/from the main grid. This possibility can be represented mathematically as,

    \begin{equation}
        {e_k^g} = \begin{cases}
        \gamma_k^{pur} \cdot P_k^g,&{\text{if}}\ P_k^g \geq 0 \\ 
        \gamma_k^{sell} \cdot P_k^g,&{\text{otherwise.}} 
        \end{cases}
    \end{equation}
where $\gamma_k^{pur}$ and $\gamma_k^{sell}$ are prices for buying/selling energy from/to the grid and $P_k^g$ is the power exchange with the grid. Following the same procedure described above, the piecewise statement can be replaced by a binary variable $\delta_k^g$ such that $P_k^g \geq 0 \Leftrightarrow \delta_k^g = 1$. This is added as a constraint to the optimization problem as in (\ref{eq:4}).
    \begin{subequations}\label{eq:4}
        \begin{align}
            -P_{min}^g \cdot \delta_k^g-P_k^g + P_{min}^g & \leq 0 \label{eq:4A}\\
            -(P_{max}^g + \varepsilon)  \cdot \delta_k^g+P_k^g+\varepsilon &\leq 0 \label{eq:4B}
        \end{align}
    \end{subequations}
where $P_{min}^g$ and $P_{max}^g$ are the minimum and maximum amount of power exchange with the grid. The resulting bi-linear term can similarly be replaced by an auxiliary variable $\psi_k^g$ such that $\psi_k^g=\delta_k^g \cdot P_k^g$. This is again included as a constraint in the optimization problem (\ref{eq:33}). Thus the interaction of the microgrid with the main grid can then be written as in (\ref{eq:45}).
    \begin{subequations}\label{eq:33}
        \begin{align}
            \psi_k^g - P_{max}^g\cdot \delta_k^g &\leq 0 \label{eq:33A}\\
            \psi_k^g - P_{min}^g\cdot \delta_k^g &\geq 0 \label{eq:33B}\\
            \psi_k^g - P_k^g + P_{min}^g(1 - \delta_k^g) &\leq 0 \label{eq:33c}\\
            \psi_k^g - P_k^g + P_{max}^g(1 - \delta_k^g) &\geq 0 \label{eq:33d}
        \end{align}
    \end{subequations}
    
    \begin{equation}
          e_k^g = (\gamma_k^{pur}-\gamma_k^{sell})\cdot \psi_k^g + \gamma_k^{sell}P_k^g
          \label{eq:45}
    \end{equation}
\subsubsection{Constraints}
The power demand of the load in the microgrid should be met at all times. This is ensured by adding a power balance constraint into the optimization problem as (\ref{eq:46}).
    \begin{equation}
          P_{k+i|k}^{pv}+P_{k+i|k}^b+P_{k+i|k}^g-P_{k+i|k}^l=0
          \label{eq:46}
    \end{equation}
where $P_{k+i|k}^{pv}$ and $P_{k+i|k}^l$ are the forecasted PV and load power respectively at instant $k$ for $i=0,1...N_p-1$. To protect the battery from rapid degradation due to overcharging and discharging, an operational limit is set on the SOC of the battery. This is given by (\ref{eq:47}).
    \begin{equation}
          SOC_{min} \leq x_k \leq SOC_{max}
          \label{eq:47}
    \end{equation}

\subsubsection{Objective Function}

The objective function that is optimized in this work can be represented as the combination of the cost functions for the grid and the storage system,
    \begin{equation}
        J = \sum_{k=1}^{N_p} (J_k^{grid} + J_k^{BESS})
    \end{equation}
where $J_k^{grid}$ and $J_k^{BESS}$ represent the cost functions for the grid and the storage system respectively. The first term in $J_k^{grid}$ minimizes the overall cost of the net energy exchange with the grid and the second term adds a penalty for every unit of energy imported from the grid. Its purpose is to discourage grid usage and to prioritize PV or battery use. On the other hand, $J_k^{BESS}$ minimizes the cost associated with the use of the battery. It penalizes both the frequency and intensity of using the battery. They are given by (\ref{eq:6A}) and (\ref{eq:6B}) respectively.
    \begin{subequations}\label{eq:6}
        \begin{align}
            J_k^{grid}  &=  e_k^g\cdot \Delta t + \lambda_{imp}\cdot \psi_k^g \cdot  \Delta t \label{eq:6A}\\
            J_k^{BESS}  &=  \alpha _{cyc}\cdot (2\psi_k^b -P_k^b)\cdot \Delta t \label{eq:6B}
        \end{align}
    \end{subequations}
where $\lambda_{imp}$ and $\alpha _{cyc}$ represent the penalty costs associated with the import from the grid and the cycling of the battery respectively. The battery cycling cost serves as a penalty to excessive cycling, a practice that can significantly degrade the overall life cycle of the battery. This parameter holds particular significance due to the stochastic nature of PV output power, variations in load, and the fluctuating spot prices over time. The penalty coefficients $\lambda_{imp}$ and $\alpha_{cye}$ in this paper were selected to balance cost minimization. Specifically, $\lambda_{imp}$ was set relative to the grid electricity price to discourage unnecessary imports, while $\alpha_{cye}$ was chosen to ensure that the optimization favors local PV utilization without incurring excessive battery cycling.

The overall optimization problem can therefore be formulated as,

    \begin{equation}
        \label{eq:14}
        \begin{aligned}
            \min \quad & \sum_{k=1}^{N_p} (J_k^{grid} + J_k^{BESS})\\
            \textrm{s.t.} \quad & \text{Battery model}\ (\ref{eq:soc})\\
            &\text{Grid model}\ (\ref{eq:4}),(\ref{eq:33}) \\
            &\text{Constraints}\ (\ref{eq:2}),(\ref{eq:3}),(\ref{eq:46}),(\ref{eq:47}) \\
        \end{aligned}
    \end{equation}
The resulting MILP problem is formulated and solved in Python using the Gurobi solver. In this study, a comprehensive analysis spanning a duration of two months is conducted to examine the system's performance under various conditions.

\section{LSTM-based PV Output Forecasting}
\label{sec:lstm}

\begin{figure}[ht]
        \centering
        \subfloat[]{\includegraphics[scale = 1.2]{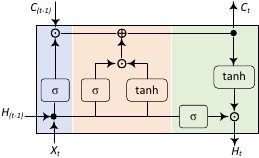}%
        \label{fig_04_a}}
        \hfil
        \subfloat[]{\includegraphics[scale = 1.1]{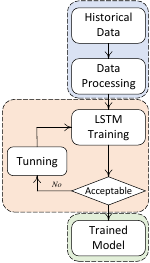}%
        \label{fig_04_b}}
        \caption{LSTM-based PV generation forecast. (a) Internal architecture of an LSTM unit (input (Xt), hidden state (Ht), and cell state (Ct). (b) Flowchart showing the steps involved in designing the LSTM-based PV generation forecast model \cite{dimd2022review}.}
        \label{fig_sim}
    \end{figure}

In this study, LSTM network was selected for developing the PV generation forecast model. LSTM is a type of Recurrent Neural Network (RNN) specifically designed to capture long-term dependencies and patterns in sequential data, such as PV generation data \cite{hochreiter1997long}. In contrast to conventional RNNs, which suffer from the vanishing gradient problem leading to insignificant weight updates, LSTM mitigates this issue by incorporating gates within its internal network structure. The choice of LSTM in this work is based on its effectiveness in a previous study \cite{dimd2022ultra}, where it provided better performance in forecasting PV output power for the same plant considered in this work (refer Section \ref{sec:pv_res}). Figure \ref{fig_04_a} and Figure \ref{fig_04_b} illustrate, respectively, the internal structure of an LSTM unit and the sequential steps involved in designing an LSTM-based PV generation forecasting model. The LSTM unit comprises a memory cell and various gates that regulate the flow of information into, out of, and within the memory cell (Figure \ref{fig_04_a}).

The LSTM-based PV generation forecast model was iteratively optimized using the validation dataset. This optimization involved tuning various parmeters and hyperparameters of the LSTM network, including the number of hidden layers, units per layer, batch size, learning rate, dropout rate, epoch size, among others. The objective was to identify the optimal combination of hyperparameters that maximizes the model's forecasting accuracy and generalization capability on a test dataset. In this study, the following optimal hyperparameter values were obtain and used: two hidden layers, each containing 10 units, a batch size of 24, 150 epochs, and a learning rate of 0.001. Additionally, since the forecast is based on time-series prediction using historical PV data, a lookback window of 96 steps (24 hours) was employed to generate predictions for the next 24 hours (96 steps).

%------------------------------------------------------------------------------------------------------------%

\section{Result and Discussion}
\label{sec:res}

This section presents the results from two key components of the study: the PV generation forecast model and the optimal operation of the microgrid. First, the performance of the LSTM-based forecasting model is assessed briefly. Then, the result of the microgrid optimization process are analyzed, focusing on cost minimization and battery usage under the forecasted PV profiles.

\subsection{PV Generation Forecast}
\label{sec:pv_res}

This section summarizes the main findings regarding the performance of an LSTM-based PV output power forecasting model for a \text{20 kWp} plant located in Trondheim, Norway. This plant is installed on the rooftop of the Department of Electric Energy at the Norwegian University of Science and Technology (NTNU). A persistence model (PM), which assumes that the forecast at time t+1 is identical to the observed value at time t, is used as the baseline for comparison. Data collected over a one-year period (2020) with a temporal resolution of 15 minutes is used. Months with incomplete data and winter months (January, February, October, November, and December) were excluded from the training dataset. From the remaining data, 153 days of data (72\%) were used for training and validation, and 61 days of data (28\%) reserved for testing. The forecast models are evaluated based on Root Mean Square Error (RMSE), Weighted Absolute Percentage Error (WAPE), and the coefficient of determination ($R^2$). RMSE measures the average magnitude of the prediction errors, and it is sensitive to large errors. WAPE expresses the total absolute error as a percentage of the total actual values, providing a scale-independent measure. The $R^2$ metric indicates how effectively the predictions capture the variability of the actual data, with a value of 1 representing a perfect fit.

Table \ref{tab:metrics} presents a statistical summary of the LSTM-based forecast model's performance for a single-step ahead (15 minutes) prediction. It is not surprising to observe that the LSTM model outperforms the PM model across all evaluated metrics. This outcome aligns with expectations since the PM model typically performs well only under stable weather conditions, which were not the case in this study. The LSTM-based forecast model achieves a performance improvement of approximately 6\% in terms of RMSE, representing a modest incremental benefit considering the model's complexity for such a short-term forecasting horizon. This suggests that for very short-term predictions, simpler methods might already provide reasonably good accuracy. Therefore, the additional computational complexity and resource requirements of the LSTM model may only be justified if better improvements can be realized at longer forecast horizons in certain application, as will be discussed in the following section.

\begin{table}[h]
    \caption{Comparison of forecasting performance between a persistence model (PM) and an LSTM-based model for single-step-ahead (15-minute) PV output power prediction based on the RMSE, WAPE and $R^2$ evaluation metrics.}
    \centering
    \begin{tabular}{rccc}
    \hline
    \hline
         & RMSE (kW) & WAPE (\%) & $R^2$\\
         \hline
        PM  & 1.036 & 23.0 & 0.88\\
        LSTM & 0.973 & 22.24 & 0.8946 \\
        \hline
        \hline
    \end{tabular}

    \label{tab:metrics}
\end{table}

\begin{figure}[ht]
        \centering
        \includegraphics[scale=0.8]{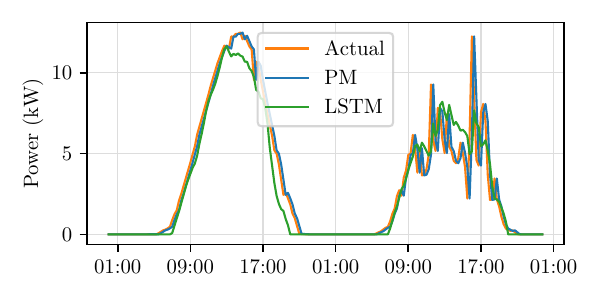}
        \caption{Comparison of actual and forecasted PV output power for two randomly selected consecutive days from the test dataset (August 10 and 11), showing a single-step-ahead (15-minute) prediction.}
        \label{fig:pv_forecast}
    \end{figure}

Figure \ref{fig:pv_forecast} illustrates the actual and predicted PV output power for two randomly selected days from the test dataset. It is evident from these figures that the PM model closely follows the actual power output for the 15-minute-ahead forecast. However, the LSTM-based model demonstrates good generalization ability, even though it struggles with rapid changes in PV power. This observation is an indication that the LSTM-based model has the potential to outperform the PM model over longer forecast horizons.

\subsection{Optimal Operation of Microgrid}

This section presents the main results from the optimal operation of the microgrid. It begins with a brief description of the microgrid control parameters and data sources employed in the optimization process. Subsequently, the evaluation metrics used to assess the performance of different forecast scenarios are defined. Finally, the behavior of the MPC-based energy scheduling strategy is analyzed under three PV generation forecast scenarios.

\subsubsection{Simulation Parameters and Data Sources}

The microgrid and simulation parameters used in this work are summarized in Table \ref{tab_01}. As discussed in the previous section, the PV output power profile is obtained from a $20\ \text{kW}_p$ PV plant located in Trondheim, Norway. The load profile used in this study is based on the average electricity consumption of 100 Norwegian households, based on smart meter measurements \cite{berg2022data}. This averaged profile represents a typical demand pattern. For the purposes of this work, the load data has been scaled to match the PV generation profile, while preserving its temporal variations. The real-time electricity market price data used in this study corresponds to the day-ahead market operated by Nord Pool, the Norwegian electricity market. This data is publicly available \cite{entsoe_prices_no3_2020}. In the optimization, the selling price of electricity to the grid was set to 75\% of the purchase price. This assumption reflects practical tariff structure in Norway, where purchase prices include grid tariffs and taxes that are not compensated when exporting PV energy. This assumption also prevent unrealistic arbitrage and encourages local consumption of PV generation. For the BESS, a capacity of $30\ \text{kWh}$ was assumed in the optimization. Such sizing allows the battery to meaningfully increase self-consumption by shifting midday surplus to evening demand, while avoiding over-sizing that would lead to low utilization during winter months.

    \begin{table}[ht]
        \caption{Microgrid and control parameters used in the optimization process.}
        \label{tab_01}
        \centering
        \small
            \begin{tabular}{l|l}
                \hline
                \hline
                 Component&Value\\
                \hline
                Grid & $P_{max}^g = 15\ kW$ and $P_{min}^g = -15\ kW$\\
                \hline
                 & $C_b=30\ kWh$, $SOC_{min}=0.2$\\
                Battery & $SOC_{max}=0.9$, $P_{max}^b = 15\ kW$\\
                  & $P_{min}^b = -15\ kW$, $\eta_{ch} = 0.9$, $\eta_{dis} = 0.95$\\
                  \hline
                PV Capacity  & $20\ kW_p$\\
                \hline
                Simulation  & $\Delta t = 15\ min$, $N_p = 24\ hrs$\\
                \hline
                \hline
            \end{tabular}
    \end{table}

\begin{figure*}[htb]
        \centering
        \includegraphics[scale=0.9]{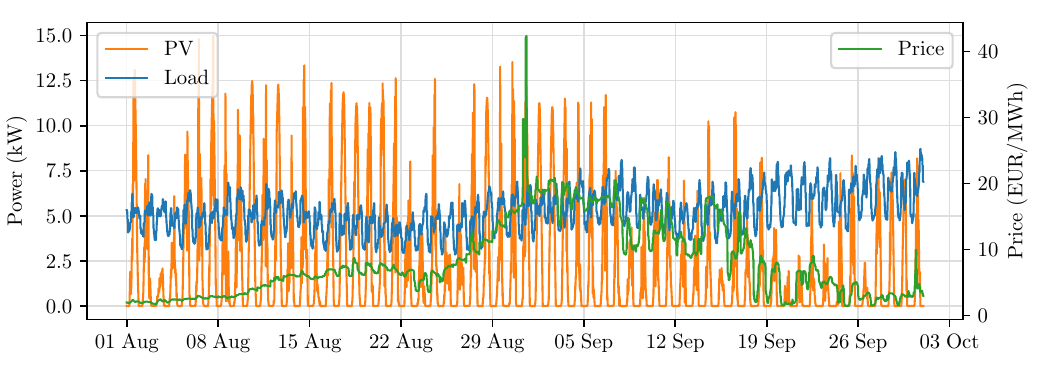}
        \caption{PV power profile, load consumption, and electricity price used for the optimal operation of the microgrid. Data covers 61 days (August 1 to September 30, 2020).}
        \label{fig:pv_load_price}
    \end{figure*}

The microgrid operation is optimized over a two-month simulation period (August to September) using a 15-minute timestep. The initial MPC operation was implemented for a horizon of 24 hours (96 steps). The optimization problem is formulated and solved in Python using the Gurobi solver. Figure \ref{fig:pv_load_price} shows the PV power, load demand, and electricity price data used in the optimization process.

\subsubsection{Evaluation Metrics}

To evaluate the impact of PV generation forecast and the control policy on the optimal operation of the grid connected PV-BESS microgrid system, three metrics are considered here:

\begin{itemize}
    \item Total energy cost: It represents the net cost of energy transactions with the grid. It accounts for electricity purchased to meet the load and subtracts revenue earned from exporting excess PV energy. It reflects the economic efficiency of the operation strategy under time-varying electricity prices. 
    \item Self-consumption ratio (SCR): It quantifies the proportion of PV energy that is consumed locally; either directly by the load or indirectly through battery storage. A higher SCR indicates more effective use of local generation and reduced dependency on the grid. 
    \item Battery cycle throughput: This measures the total energy charged and discharged through the battery during the control period. It indicates the intensity and frequency of battery usage.
\end{itemize}

\subsubsection{Analysis of the MPC operation}

To evaluate the impact of PV generation forecast accuracy on the optimal operation of the microgrid, three forecasting scenarios are considered within the MPC framework. In the first scenario, referred to as the ideal case, the MPC has perfect knowledge of PV generation over the control horizon, representing the upper bound of system performance under perfect information. The second scenario, representing a simplistic case, uses a naive persistence model where the PV output is assumed to remain constant based on the most recent observation, simulating a situation with poor forecast accuracy. The third scenario reflects a practical implementation in which the MPC utilizes PV forecasts generated by an LSTM neural network model, as described in the previous section. It is important to note that it is not in the scope of this work to develop a forecasting model for load demand. It is assumed that the MPC has a perfect knowledge of future load demand in all the three scenarios. This assumption is made to isolate and examine the specific impact of PV forecast quality on the economic and operational performance of the microgrid under MPC-based control.

Table \ref{tab:mpc_res} summarizes the aggregate energy exchanges, economic costs, SCR, and battery cycling over a two month MPC simulation for three PV forecast scenarios. Under the perfect forecast scenario (Case 1), the controller minimizes grid dependence by importing 4937 kWh and exporting only 139 kWh. This results in the lowest grid cost (\euro 37.25), the highest SCR (90.9\%), and a battery throughput of 1182 kWh. This level of battery throughput indicates that the battery is being effectively utilized to maximize self-consumption and minimize both grid interaction and dependency.

\begin{table}[h]
    \caption{Summary of grid imports, exports and associated costs, SCR, and battery throughput over the entire optimization period with a prediction horizon of 96 steps (24 hours).}
    \centering
    \begin{tabular}{cccccc}
    \hline
    \hline
         & Import & Export &Grid cost & SCR & Battery\\
        Case & (kWh) & (kWh) & (EUR) & (\%) & (kWh)\\
        \hline
         1 &  4937  & 139  & 37.25 & 90.9 & 1182\\
         2 &  5362  & 623  & 38.03 & 78.1 & 419\\
         3 &  5141  & 111  & 40.43 & 84.5 & 747\\
    \hline
    \hline
    \end{tabular}
    \label{tab:mpc_res}
\end{table}

In contrast, the second scenario (Case 2), based on the persistence model, incurs the highest grid imports (5362 kWh) and exports (623 kWh), reflecting significant misalignment between the PV forecast and actual PV generation over the control horizon. As a results it relies more on grid import to satisfy the load. Consequently, the SCR drops to 78.1\%, and battery is severely underutilized with only 419 kWh of throughput. Coincidentally, the excessive imports and exports roughly counterbalance each other, resulting in only a slightly higher cost (\euro 38.03) compared to Case 1. However, the system remains highly dependent on the grid, and the large exports could potentially contribute to grid congestion and other operational challenges.

The LSTM-based forecast case (Case 3) demonstrates intermediate performance. It imports 5141 kWh and exports 111 kWh, showing a substantially lower grid export than the persistence case. Case 3 achieves an SCR of 84.5\% and a battery throughput of 747 kWh, indicating improved utilization of both PV and storage system. However, it incurs the highest total cost (\euro 40.43), suggesting that residual forecast errors, although smaller than in the persistence model, still lead to suboptimal charge/discharge timing, particularly around grid price peaks.

\begin{figure*}[htb]
    \centering
    \subfloat[Case 1: Perfect forecast]{%
        \includegraphics[scale=0.48]{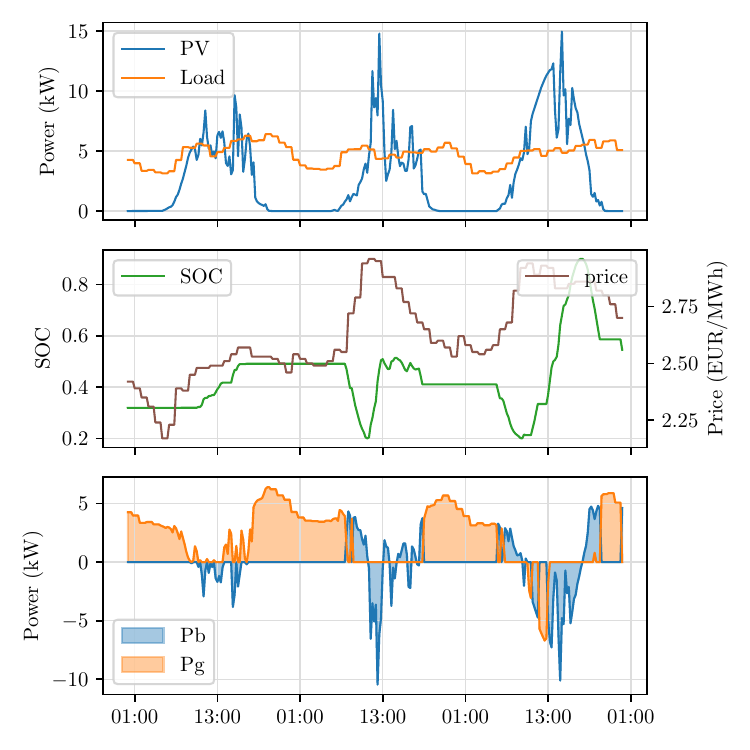}%
        \label{vvv}}
    \hfil
    \subfloat[Case 2: Persistence forecast]{%
        \includegraphics[scale=0.48]{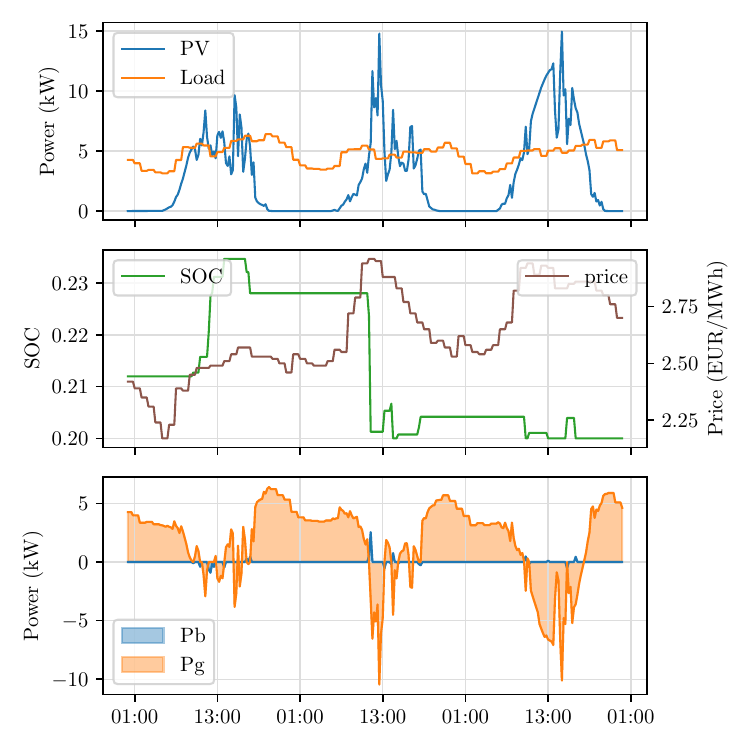}%
        \label{yyy}}
    \hfil
    \subfloat[Case 3: LSTM-based forecast]{%
        \includegraphics[scale=0.48]{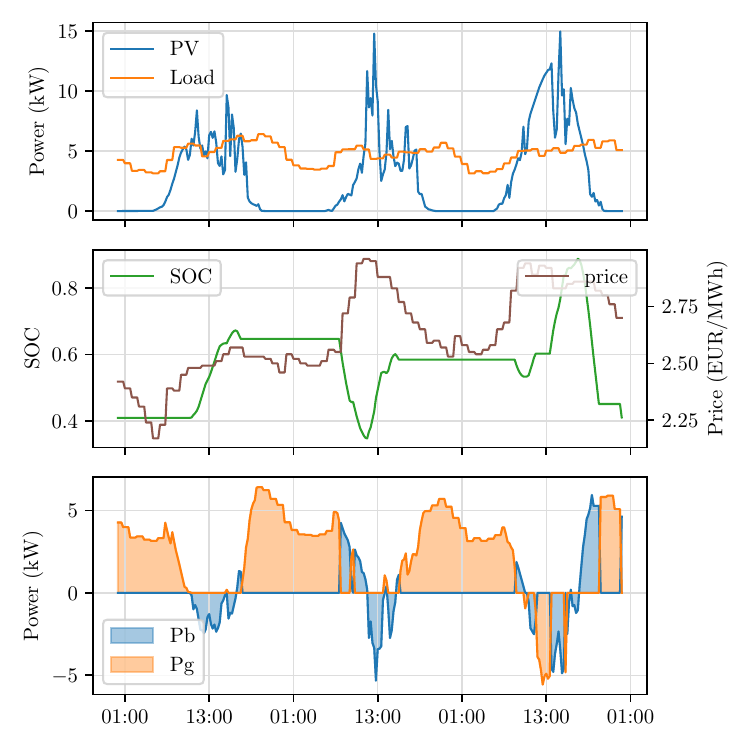}%
        \label{uuu}}
    \caption{Microgrid operation during high PV surplus days (August 5--7) under three PV forecast scenarios. Top panel: PV production and load demand. Middle panel: battery SOC and energy price. Bottom panel: battery power ($P_b$) and grid power ($P_g$).}
    \label{fig:surplus}
\end{figure*}

\begin{figure*}[htb]
        \centering
        \subfloat[Case 1: Perfect forecast]{%
            \includegraphics[scale=0.48]{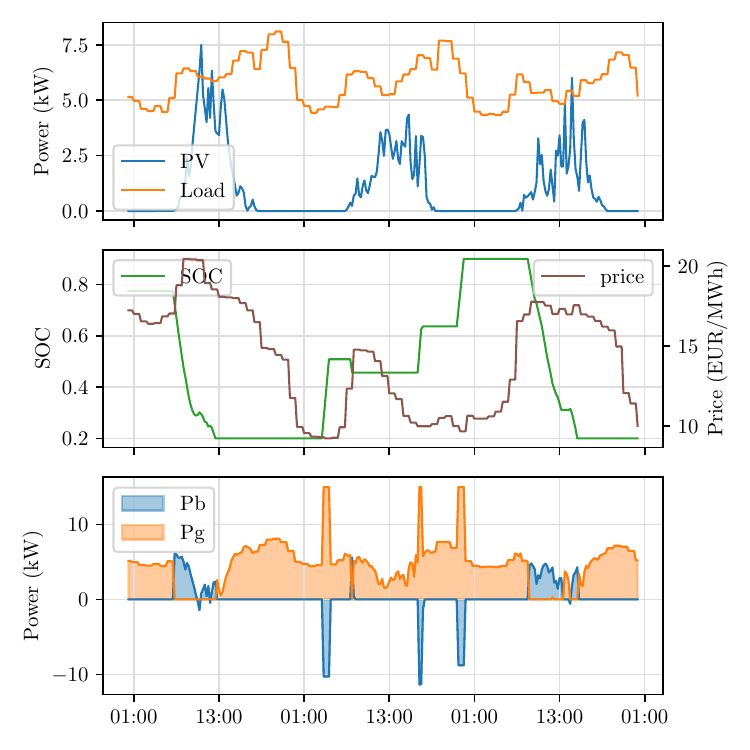}%
            \label{vv}}
        \hfil
        \subfloat[Case 2: Persistence forecast]{%
            \includegraphics[scale=0.48]{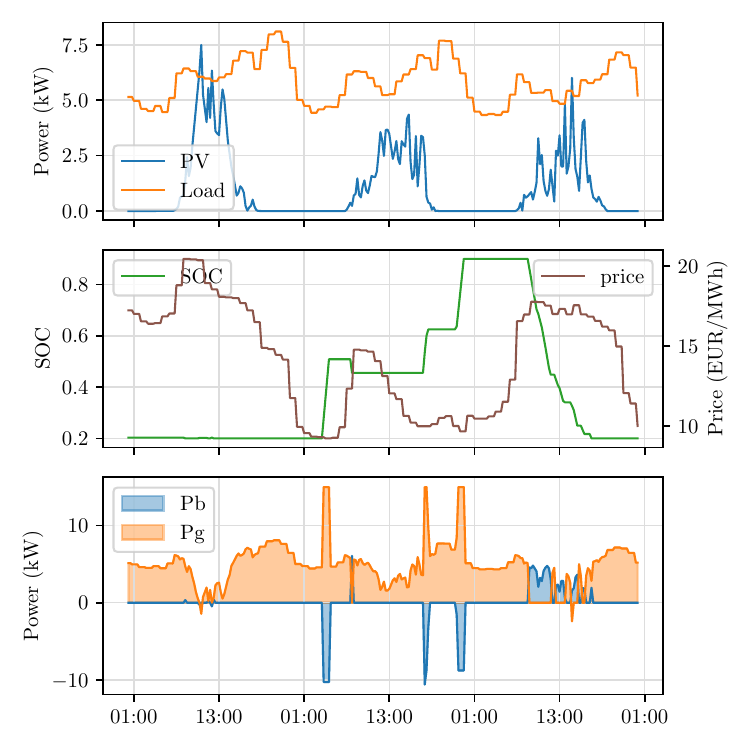}%
            \label{yy}}
        \hfil
        \subfloat[Case 3: LSTM-based forecast]{%
            \includegraphics[scale=0.48]{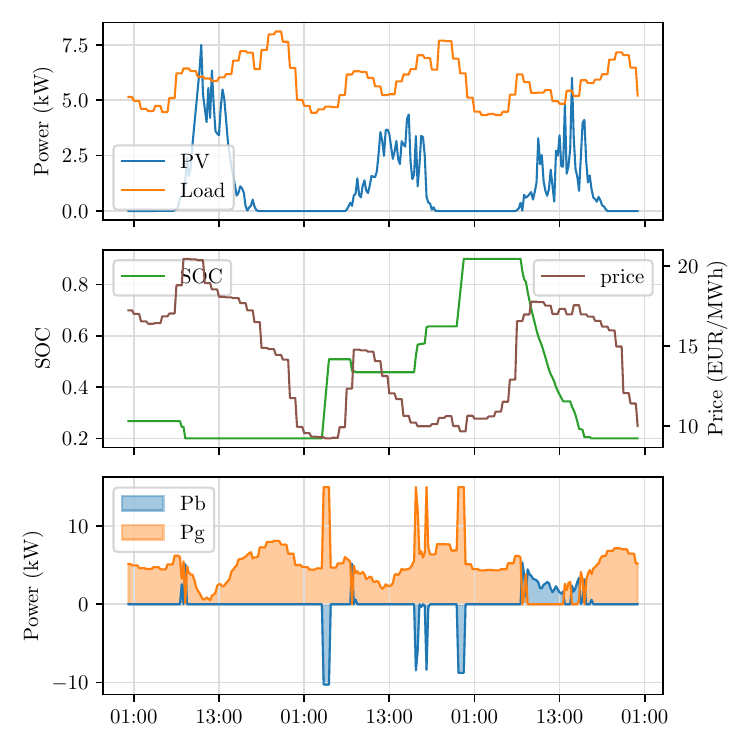}%
            \label{uu}}
        \caption{Microgrid operation during very low PV surplus days (September 7–9) under three PV forecast scenarios. Top panel: PV production and load demand. Middle panel: battery SOC and energy price. Bottom panel: battery power ($P_b$) and grid power ($P_g$).}
        \label{fig:no_surplus}
    \end{figure*}

To further evaluate the performance of the MPC-based optimization under the three forecast scenarios, two representative days, one with a `high PV surplus' and another with a `very low PV surplus', are selected. The system’s behavior on these days is illustrated in Figures \ref{fig:surplus} and \ref{fig:no_surplus}. The top panel of these figures shows the PV generation profile (PV) and load demand (Load) for the selected days, while the electricity price is shown for the same period in the middle panel. These profiles remain identical across all forecast scenarios. The battery state of charge (SOC) is shown in the middle panel, and the power flow through the battery ($P_b$) together with the power exchange with the grid ($P_g$) are shown in the bottom panel. Negative values of $P_b$ and $P_g$ indicate power flowing into the battery (charging) and export to the grid, respectively.

On high PV surplus days (Figure \ref{fig:surplus}), the perfect forecast scenario accurately anticipates PV generation peaks, enabling proactive battery charging during surplus PV hours and discharging during evening hours and price spikes. The state of charge (SOC) follows a smooth ramp-up to near full capacity before the afternoon peak prices, followed by a discharge phase. This behavior results in higher battery throughput, increased local consumption, negligible energy exported to the grid, and the lowest total grid cost. Under the persistence forecast scenario, the controller lacks foresight into future PV availability and charges the battery only modestly during the surplus period, leaving significant PV energy unused by the storage system. As a result, the SOC remains well below full capacity (between 0.2 and 0.24) throughout the day, and the controller ultimately relies more on grid imports during later price spikes. This conservative charging strategy limits both battery throughput and self-consumption, as reflected by lower battery cycling, higher energy exports, and a reduced SCR (as shown in Table \ref{tab:mpc_res}). The LSTM forecast-based controller predicts PV peaks and availability with sufficient certainty to manage battery charging and discharging more effectively. The SOC in this case follows a pattern similar to the perfect forecast scenario, discharging during peak price periods and minimizing exports. However, minor timing errors in charging during the largest PV surpluses still occur.

In the low PV surplus scenario (Figure \ref{fig:no_surplus}), PV generation is insufficient to meet the load for extended periods. Across all three cases, the MPC's SOC evolution becomes driven primarily by price signals rather than the PV availability forecast when scheduling battery operations. This results in a similar SOC pattern across all three forecast scenarios, particularly during the second and third days. However, the perfect and LSTM-based forecast scenarios show slightly better performance in utilizing the storage system. The figure also shows that, in all three cases, the battery charges from the grid during low-price hours (indicated by the large import peaks) and discharges during high-price hours.

\begin{figure}[H]
        \centering
        \includegraphics[scale=0.75]{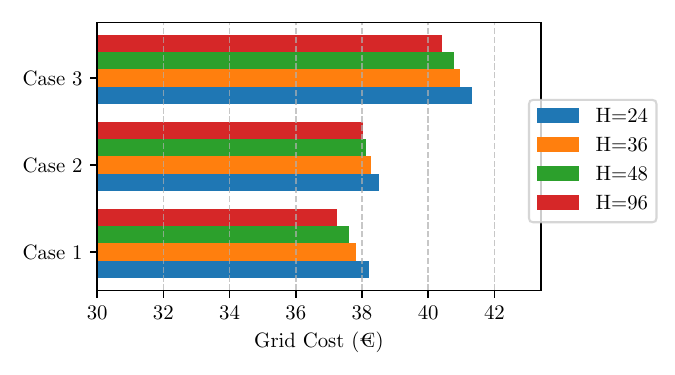}
        \caption{Effect of prediction horizon on grid cost under different PV forecast scenarios.}
        \label{fig:cost_horizon}
    \end{figure}

\begin{figure}[H]
        \centering
        \includegraphics[scale=0.75]{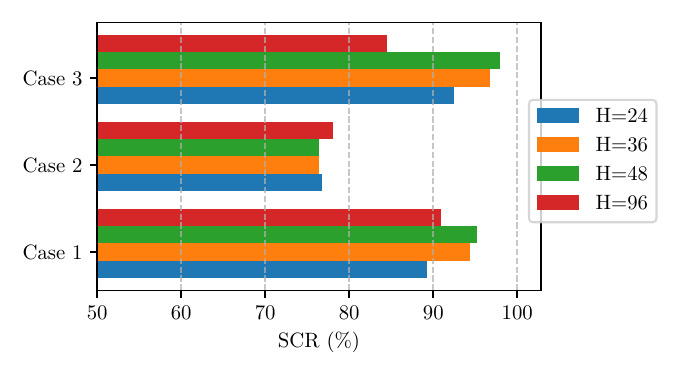}
        \caption{Effect of prediction horizon on self-consumption ratio under different PV forecast scenarios.}
        \label{fig:scr_horizon}
    \end{figure}

\begin{figure}[H]
        \centering
        \includegraphics[scale=0.75]{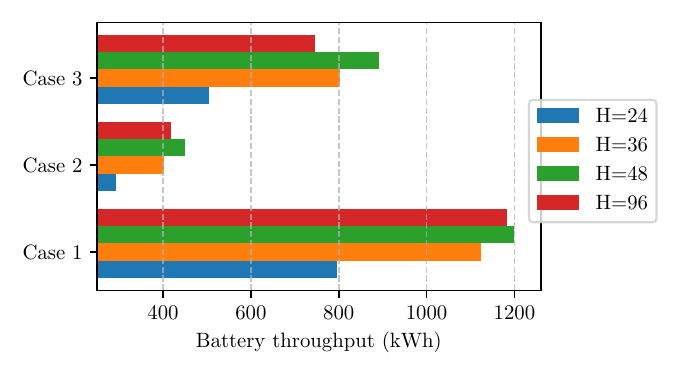}
        \caption{Effect of prediction horizon on battery energy throughput under different PV forecast scenarios.}
        \label{fig:batt_horizon}
    \end{figure}

To further assess the impact of PV forecast accuracy on the optimal operation of a microgrid under an MPC framework, the three forecast scenarios were evaluated across four prediction horizons (H = 24, 36, 48, and 96). This analysis examines how the length of the forecast window influences operational performance. The results, focusing on three performance indicators, net grid cost, SCR, and battery throughput, are presented in Figures \ref{fig:cost_horizon}, \ref{fig:scr_horizon}, and \ref{fig:batt_horizon}, respectively. As shown in Figure \ref{fig:cost_horizon}, grid cost generally decreases with increasing prediction horizon across all forecast scenarios, particularly in the ideal (Case 1) and LSTM-based (Case 3) cases. This trend is attributed to the improved accuracy of PV generation forecasts, which allows more effective control actions for optimizing the timing of imports and exports. In contrast, the persistence model (Case 2) exhibits limited responsiveness to longer horizons, with grid costs remaining relatively flat.

Figure \ref{fig:scr_horizon} shows that SCR increases with longer prediction horizons in Case 1 and Case 3, peaking at H=48 (95.2\% and 97.9\% respectively), where the controller has sufficient foresight to store excess PV energy effectively. Beyond this point (H=96), SCR begins to decline, particularly in the LSTM scenario, suggesting reduced forecast accuracy at extended horizons. In contrast, SCR in the persistence case remains consistently low and flat across all horizons due to limited forecasting capability.

Finally, Figure \ref{fig:batt_horizon} shows the effect of prediction horizon on the energy throughput of the battery storage system. As shown in the figure, battery throughput increases with horizon length in both the ideal and LSTM scenarios, indicating more effective energy shifting due to better PV generation forecasts. Similar to the trend observed in SCR, peak throughput occurs at H=48, reaching 1200 kWh in Case 1 and 892 kWh in Case 3, after which it slightly declines, likely due to increased forecast uncertainty and more conservative battery operation. In contrast, battery throughput in the persistence scenario remains consistently low and largely unaffected by horizon length, reflecting underutilization caused by poor forecast reliability.

%------------------------------------------------------------------------------------------------------------%

\section{Conclusion}
\label{sec:conc}

Grid-connected PV systems introduce challenges to the conventional grid due to their sensitivity to environmental conditions, which causes uncertain and irregular output power. To address this variability, combining energy storage, PV output forecasting, and optimal energy scheduling in microgrids is essential. However, the effectiveness of such systems ultimately depends on the accuracy of the PV forecasting model and the design of the control policy. This study proposed and evaluated a multi-objective energy scheduling strategy for a grid-connected PV-BESS microgrid that integrates LSTM-based PV output forecasting and MPC-based control. The developed MPC-based framework enables optimal energy exchange within the microgrid and with the grid by balancing competing objectives: minimizing energy costs, increasing PV self-consumption, mitigating BESS degradation, and reducing grid stress due to PV injection. The main contribution of this work lies in analyzing the operational and economic impacts of PV forecast accuracy on microgrid performance. Through a detailed case study involving real-world PV generation, load demand, and electricity price data, the proposed framework was evaluated under three PV forecast scenarios: perfect forecast, persistence model, and LSTM-based forecast.

Results showed that PV forecast accuracy significantly affects microgrid operation. While the perfect forecast scenario achieved the best performance across all metrics, the LSTM-based approach delivered substantial improvements over the persistence model, particularly in terms of reducing large PV injections into the grid, effective battery utilization and improving PV self-consumption. It is important to note that the LSTM-based approach has higher battery throughput due to frequent cycling, which inevitably contributes to accelerated aging. However, it gives significantly better operational outcomes. By contrast, the persistence case results in low throughput, with the battery remaining idle at low SOC for long periods. While this lowers cycling-related degradation and appears favorable from the objective function’s perspective, which penalizes battery cycling, it leads to higher dependence on grid imports and substantial PV injection into the grid. The latter is undesirable, as large injections of surplus PV into the grid can result in congestion and compromise grid stability, particularly in distribution networks with high PV penetration. Thus, although the high-throughput case can negatively affects battery lifetime, it has overall better performance. Additionally, the analysis indicated that a prediction horizon of 12 hours (48 steps) gives the best trade-off between forecast accuracy and operational performance.

These findings demonstrate the importance of integrating accurate PV forecasting into predictive control frameworks for efficient microgrid operation, especially under real-time pricing. They also highlight the need for control policies that are robust to forecast uncertainty and account for battery aging and grid constraints. This study assumes perfect load forecasts and includes some simplified system dynamics (e.g., fixed battery efficiency and ideal grid access), which may not fully reflect real-world conditions. Future research could focus both on PV and load forecasting, probabilistic MPC strategies under uncertainty, and the integration of ancillary services such as demand response or grid support functionalities.

\section*{CRediT authorship contribution statement}

Berhane Darsene Dimd: Writing – original draft, Visualization, Software, Methodology, Conceptualization. Steve Völler: Writing – review and editing, Validation, Supervision. Ole-Morten Midtgård: Supervision, Conceptualization, and Funding acquisition.

\section*{Declaration of Generative AI and AI-assisted technologies in the writing process}

During the preparation of this work the authors used ChatGPT in order to improve the language and readability. After using this tool/service, the authors reviewed and edited the content as needed and take full responsibility for the content of the publication.

\section*{Declaration of competing interest}

The authors declare that they have no known competing financial interests or personal relationships that could have appeared to influence the work reported in this paper.

\section*{Data availability}

Data will be made available on request.

\section*{Acknowledgements}
This work was supported by the Department of Electric Energy, Norwegian University of Science and Technology (NTNU), Trondheim, Norway.

 \bibliographystyle{elsarticle-num} 
 \bibliography{refs}

\begin{thebibliography}{10}
\expandafter\ifx\csname url\endcsname\relax
  \def\url#1{\texttt{#1}}\fi
\expandafter\ifx\csname urlprefix\endcsname\relax\def\urlprefix{URL }\fi
\expandafter\ifx\csname href\endcsname\relax
  \def\href#1#2{#2} \def\path#1{#1}\fi

\bibitem{IEA2024electricity}
{International Energy Agency}, \href{https://www.iea.org/reports/electricity-2024}{Electricity 2024: Analysis and forecast to 2026}, Tech. rep., International Energy Agency, Paris (2024).
\newline\urlprefix\url{https://www.iea.org/reports/electricity-2024}

\bibitem{adefarati2019reliability}
T.~Adefarati, R.~C. Bansal, Reliability, economic and environmental analysis of a microgrid system in the presence of renewable energy resources, Applied energy 236 (2019) 1089--1114.
\newblock \href {https://doi.org/10.1016/j.apenergy.2018.12.050} {\path{doi:10.1016/j.apenergy.2018.12.050}}.

\bibitem{gandhi2024value}
O.~Gandhi, W.~Zhang, D.~S. Kumar, C.~D. Rodr{\'\i}guez-Gallegos, G.~M. Yagli, D.~Yang, T.~Reindl, D.~Srinivasan, The value of solar forecasts and the cost of their errors: A review, Renewable and Sustainable Energy Reviews 189 (2024) 113915.
\newblock \href {https://doi.org/10.1016/j.rser.2023.113915} {\path{doi:10.1016/j.rser.2023.113915}}.

\bibitem{bertsch2017drives}
V.~Bertsch, J.~Geldermann, T.~L{\"u}hn, What drives the profitability of household pv investments, self-consumption and self-sufficiency?, Applied Energy 204 (2017) 1--15.
\newblock \href {https://doi.org/10.1016/j.apenergy.2017.06.055} {\path{doi:10.1016/j.apenergy.2017.06.055}}.

\bibitem{hassan2017optimal}
A.~S. Hassan, L.~Cipcigan, N.~Jenkins, Optimal battery storage operation for pv systems with tariff incentives, Applied Energy 203 (2017) 422--441.
\newblock \href {https://doi.org/10.1016/j.apenergy.2017.06.043} {\path{doi:10.1016/j.apenergy.2017.06.043}}.

\bibitem{xu2016modeling}
B.~Xu, A.~Oudalov, A.~Ulbig, G.~Andersson, D.~S. Kirschen, Modeling of lithium-ion battery degradation for cell life assessment, IEEE transactions on smart grid 9~(2) (2016) 1131--1140.
\newblock \href {https://doi.org/10.1109/TSG.2016.2578950} {\path{doi:10.1109/TSG.2016.2578950}}.

\bibitem{alam2016cycle}
M.~Alam, T.~Saha, Cycle-life degradation assessment of battery energy storage systems caused by solar pv variability, in: 2016 IEEE Power and Energy Society General Meeting (PESGM), IEEE, 2016, pp. 1--5.
\newblock \href {https://doi.org/10.1109/PESGM.2016.7741532} {\path{doi:10.1109/PESGM.2016.7741532}}.

\bibitem{tevar2019influence}
G.~T{\'e}var, A.~G{\'o}mez-Exp{\'o}sito, A.~Arcos-Vargas, M.~Rodr{\'\i}guez-Monta{\~n}{\'e}s, Influence of rooftop pv generation on net demand, losses and network congestions: A case study, International Journal of Electrical Power \& Energy Systems 106 (2019) 68--86.
\newblock \href {https://doi.org/10.1016/j.ijepes.2018.09.013} {\path{doi:10.1016/j.ijepes.2018.09.013}}.

\bibitem{dhlamini2018solar}
N.~Dhlamini, S.~D. Chowdhury, Solar photovoltaic generation and its integration impact on the existing power grid, in: 2018 IEEE PES/IAS PowerAfrica, IEEE, 2018, pp. 710--715.
\newblock \href {https://doi.org/10.1109/PowerAfrica.2018.8521003} {\path{doi:10.1109/PowerAfrica.2018.8521003}}.

\bibitem{rahman2021analysis}
S.~Rahman, S.~Saha, S.~N. Islam, M.~T. Arif, M.~Mosadeghy, M.~Haque, A.~M. Oo, Analysis of power grid voltage stability with high penetration of solar pv systems, IEEE Transactions on Industry Applications 57~(3) (2021) 2245--2257.
\newblock \href {https://doi.org/10.1109/TIA.2021.3066326} {\path{doi:10.1109/TIA.2021.3066326}}.

\bibitem{bordons2020basic}
C.~Bordons, F.~Garcia-Torres, M.~A. Ridao, C.~Bordons, F.~Garcia-Torres, M.~A. Ridao, Basic energy management systems in microgrids, Model Predictive Control of Microgrids (2020) 77--107\href {https://doi.org/10.1007/978-3-030-24570-2_4} {\path{doi:10.1007/978-3-030-24570-2_4}}.

\bibitem{hao2020power}
Y.~Hao, L.~Dong, J.~Liang, X.~Liao, L.~Wang, L.~Shi, Power forecasting-based coordination dispatch of pv power generation and electric vehicles charging in microgrid, Renewable Energy 155 (2020) 1191--1210.
\newblock \href {https://doi.org/10.1016/j.renene.2020.03.169} {\path{doi:10.1016/j.renene.2020.03.169}}.

\bibitem{nwulu2017optimal}
N.~I. Nwulu, X.~Xia, Optimal dispatch for a microgrid incorporating renewables and demand response, Renewable energy 101 (2017) 16--28.
\newblock \href {https://doi.org/10.1016/j.renene.2016.08.026} {\path{doi:10.1016/j.renene.2016.08.026}}.

\bibitem{nair2021analysis}
U.~R. Nair, M.~Sandelic, A.~Sangwongwanich, T.~Dragi{\v{c}}evi{\'c}, R.~Costa-Castello, F.~Blaabjerg, An analysis of multi objective energy scheduling in pv-bess system under prediction uncertainty, IEEE Transactions on Energy Conversion 36~(3) (2021) 2276--2286.
\newblock \href {https://doi.org/10.1109/TEC.2021.3055453} {\path{doi:10.1109/TEC.2021.3055453}}.

\bibitem{deng2022economic}
X.~Deng, Z.~Deng, Z.~Song, X.~Lin, X.~Hu, Economic control for a residential photovoltaic-battery system by combining stochastic model predictive control and improved correction strategy, Journal of Energy Resources Technology 144~(5) (2022).
\newblock \href {https://doi.org/10.1115/1.4051735} {\path{doi:10.1115/1.4051735}}.

\bibitem{bordons2020energy}
C.~Bordons, F.~Garcia-Torres, M.~A. Ridao, C.~Bordons, F.~Garcia-Torres, M.~A. Ridao, Energy management with economic and operation criteria, Model Predictive Control of Microgrids (2020) 109--145\href {https://doi.org/10.1007/978-3-030-24570-2_5} {\path{doi:10.1007/978-3-030-24570-2_5}}.

\bibitem{nunez2017optimal}
A.~N{\'u}{\~n}ez-Reyes, D.~M. Rodr{\'\i}guez, C.~B. Alba, M.~{\'A}.~R. Carlini, Optimal scheduling of grid-connected pv plants with energy storage for integration in the electricity market, Solar Energy 144 (2017) 502--516.
\newblock \href {https://doi.org/10.1016/j.solener.2016.12.034} {\path{doi:10.1016/j.solener.2016.12.034}}.

\bibitem{garcia2015optimal}
F.~Garcia-Torres, C.~Bordons, Optimal economical schedule of hydrogen-based microgrids with hybrid storage using model predictive control, IEEE Transactions on Industrial Electronics 62~(8) (2015) 5195--5207.
\newblock \href {https://doi.org/10.1109/TIE.2015.2412524} {\path{doi:10.1109/TIE.2015.2412524}}.

\bibitem{conte2020economic}
E.~Conte, P.~R. Mendes, J.~E. Normey-Rico, Economic management based on hybrid mpc for microgrids: A brazilian energy market solution, Energies 13~(13) (2020) 3508.
\newblock \href {https://doi.org/10.3390/en13133508} {\path{doi:10.3390/en13133508}}.

\bibitem{np}
Day-ahead prices, \url{https://www.nordpoolgroup.com/en/}, accessed: 2025-08-03.

\bibitem{seborg2016process}
D.~E. Seborg, T.~F. Edgar, D.~A. Mellichamp, F.~J. Doyle~III, Process dynamics and control, John Wiley \& Sons, 2016.

\bibitem{stefanopoulou2015system}
A.~Stefanopoulou, Y.~Kim, System-level management of rechargeable lithium-ion batteries, Rechargeable Lithium Batteries (2015) 281--302\href {https://doi.org/10.1016/B978-1-78242-090-3.00010-9} {\path{doi:10.1016/B978-1-78242-090-3.00010-9}}.

\bibitem{bemporad1999control}
A.~Bemporad, M.~Morari, Control of systems integrating logic, dynamics, and constraints, Automatica 35~(3) (1999) 407--427.
\newblock \href {https://doi.org/10.1016/S0005-1098(98)00178-2} {\path{doi:10.1016/S0005-1098(98)00178-2}}.

\bibitem{hochreiter1997long}
S.~Hochreiter, J.~Schmidhuber, Long short-term memory, Neural computation 9~(8) (1997) 1735--1780.
\newblock \href {https://doi.org/10.1162/neco.1997.9.8.1735} {\path{doi:10.1162/neco.1997.9.8.1735}}.

\bibitem{dimd2022ultra}
B.~D. Dimd, S.~V{\"o}ller, O.-M. Midtg{\aa}rd, T.~M. Zenebe, Ultra-short-term photovoltaic output power forecasting using deep learning algorithms, in: 2022 IEEE 21st Mediterranean Electrotechnical Conference (MELECON), IEEE, 2022, pp. 837--842.
\newblock \href {https://doi.org/10.1109/MELECON53508.2022.9843113} {\path{doi:10.1109/MELECON53508.2022.9843113}}.

\bibitem{dimd2022review}
B.~D. Dimd, S.~V{\"o}ller, U.~Cali, O.-M. Midtg{\aa}rd, A review of machine learning-based photovoltaic output power forecasting: Nordic context, IEEE Access 10 (2022) 26404--26425.
\newblock \href {https://doi.org/10.1109/ACCESS.2022.3156942} {\path{doi:10.1109/ACCESS.2022.3156942}}.

\bibitem{berg2022data}
K.~Berg, M.~L{\"o}schenbrand, A data set of a norwegian energy community, Data in Brief 40 (2022) 107683.
\newblock \href {https://doi.org/10.1016/j.dib.2021.107683} {\path{doi:10.1016/j.dib.2021.107683}}.

\bibitem{entsoe_prices_no3_2020}
{ENTSO-E Transparency Platform}, \href{https://transparency.entsoe.eu/}{Day-ahead energy prices for norway no3}, accessed: 2025-08-03 (2020).
\newline\urlprefix\url{https://transparency.entsoe.eu/}

\end{thebibliography}

\end{document}